# Probabilities and certainties within a causally symmetric model

Roderick I. Sutherland

Centre for Time, University of Sydney, NSW 2006 Australia

rod.sutherland@sydney.edu.au

**Abstract:**

This paper is concerned with the causally symmetric version of the familiar de Broglie-Bohm interpretation, this version allowing the spacelike nonlocality and the configuration space ontology of the original model to be avoided via the addition of retrocausality. Two different features of this alternative formulation are considered here. With regard to probabilities, it is shown that the model provides a derivation of the Born rule identical to that in Bohm's original formulation. This derivation holds just as well for a many-particle, entangled state as for a single particle. With regard to "certainties", the description of a particle's spin is examined within the model and it is seen that a statistical description is no longer necessary once final boundary conditions are specified in addition to the usual initial state, with the particle then possessing a definite (but hidden) value for every spin component at intermediate times. These values are consistent with being the components of a single, underlying spin vector. The case of a two-particle entangled spin state is also examined and it is found that, due to the retrocausal aspect, each particle possesses its own definite spin during the entanglement, independent of the other particle. In formulating this picture, it is demonstrated how such a realistic model can preserve Lorentz invariance in the face of Bell's theorem and avoid the need for a preferred reference frame.

## 1. Introduction

The de Broglie-Bohm model [1-3] is notable in providing a simple physical picture that could underlie the mathematics of quantum mechanics. Unfortunately, it is also notable that this otherwise appealing ontology requires a preferred reference frame in order to accommodate the nonlocality implied by Bell's theorem [4], thereby conflicting with relativity. In response to this fact, the present author put forward a retrocausal generalisation of the de Broglie-Bohm model some time ago [5,6] in order to show that this violation of Lorentz invariance at the hidden level is not essential and can be avoided by relaxing a common presumption about causality. This alternative essentially involves assuming that a final boundary condition needs to be specified in addition to the usual initial one in order to determine all intervening values fully. In formulating this possible refinement, however, the main focus has previously been on describing the details of the underlying reality it implies, not on providing a corresponding derivation for the known probabilities of quantum mechanics.

A pleasing feature of the standard de Broglie-Bohm model is Bohm's theory of measurement [2] in which the Born probability rule for a measurement of any observable (i.e., not just position) can be derived once the initial distribution for particle positions is assumed to have the familiar $|\psi|^2$ form. By contrast, the retrocausal version, which was dubbed the "causally symmetric" model, merely introduced the Born rule as a separate postulate, without any attempt at a derivation. This left the model open to the criticism that it was less effective because it could not explain the observed probabilities to the same extent that the original model could.

The first aim of the present paper is to point out that this criticism is unwarranted and that the measurement theory of the original model can actually be carried across unchanged to the causally symmetric version. (The present author originally overlooked this point when formulating the alternative model.) This resource is possible because, although the two versions predict different particle trajectories, both employ the same probability distribution for particle positions in the circumstance where only the usual, initial wavefunction is given. This feature is all that is required for the derivation to remain valid, as will be discussed here in more detail.

The second aim of the present paper is to examine the different ontology and the hidden values which become apparent once a final boundary condition is introduced. The discussion will focus on spin as a convenient observable for illustrating the differences. In addition to accommodating Lorentz invariance, avoiding spacelike influences and generally providing a more time-symmetric picture, the causally symmetric description has the agreeable characteristic of allowing the ontology for n entangled particles to remain in spacetime rather than being banished to 3n-dimensional configuration space. The usual correlated probabilities are still correctly predicted because the original theory of measurement is maintained.

## 2. Comparison of the two models

In this section only the single-particle case will be considered since this is sufficient to illustrate the essential ideas. The entangled, many-particle case will be addressed in later sections.

Both the standard and causally symmetric versions of the de Broglie-Bohm model start by assuming that a particle has a definite trajectory. The standard version then proceeds via the Schrodinger current density[1]:

$$\mathbf{j} = \frac{\hbar}{2im}\psi^* \overleftrightarrow{\nabla} \psi \qquad (1)$$

and takes the hidden velocity of the particle to be in the same direction as this current density. Here the wavefunction $\psi$ is the result of prior preparation, e.g., the outcome of a previous measurement, and can be viewed as the initial boundary condition imposed. This state, together with the hidden initial position of the particle, then provides a deterministic description of the particle's future motion. In contrast, the basic idea of the causally symmetric model is that the initial state is not sufficient and that a final boundary condition is also required in order to determine the trajectory of the particle at intermediate times. By symmetry, this final condition is taken to be represented by a separate and independent $\psi$. It can most simply be viewed as the result of the next measurement performed. From here on, the initial and final wavefunctions will be distinguished by writing them as $\psi_i$ and $\psi_f$, respectively. In this notation, the current density expression used for the causally symmetric version is[2]:

$$\mathbf{j} = \mathrm{Re}\frac{\hbar}{2ima}\psi_f^* \overleftrightarrow{\nabla} \psi_i \tag{2}$$

where a is a normalisation constant consisting of the inner product of $\psi_i$ and $\psi_f$. This expression differs from the usual one in Eq. (1) in that both initial and final boundary conditions are included and that the real part has been taken for the ontology to be consistent with real spacetime. As in the standard de Broglie-Bohm model, the hidden velocity of the particle is taken to be in the same direction as the current density. Both current density expressions satisfy equations of continuity derivable from the Schrodinger equation. This ensures local conservation of probability and compatibility with continuous trajectories. Since a final boundary condition, such as the result of the next measurement, cannot be known in advance, standard quantum mechanics arises from this model from the need to take a weighted average over the unknown final possibilities. It is readily shown that Eq. (2) reduces back to Eq. (1) under these circumstances (see Appendix).

[1] Here m is the particle's mass, $\hbar$ is Planck's constant, $i = \sqrt{-1}$, $\psi(\mathbf{x};t)$ is the particle's wavefunction, the operator $\overleftrightarrow{\nabla}$ is an abbreviation for $\overrightarrow{\nabla} - \overleftarrow{\nabla}$ and the grad operators $\overrightarrow{\nabla}$ and $\overleftarrow{\nabla}$ act to the right and left, respectively.

## 3. Retrocausality defined

A short digression will now be made to explain further what is meant by retrocausality[3]. The uncertainty principle of quantum mechanics highlights two important facts, namely (i) that there are always some quantities whose future values are left open and cannot be predicted based on previous measurement results, and (ii) that measurements not only provide information on a particle's state but also exert an influence which can change that state as well. As a result of the first point, the situation differs from classical mechanics in that there is room for the experimenter's choice for the next observable to be measured to have an effect on earlier values that were left unknown by the previous measurement. This would be a backwards-in-time, or "retrocausal", effect. This possibility is reinforced by the second point that measurements can exert an influence as well as providing information. Hence the concept being introduced here is that the choice of measurement interaction not only affects the subsequent situation but also affects the prior situation. This is described mathematically by the inclusion of the future measurement outcome $\psi_f$ in Eq. (2) governing the particle's present trajectory[4]. Stated more precisely, the experimenter's role is to choose the set of allowable eigenstates which could arise from a measurement, thereby influencing and restricting the possibilities available for the particle's hidden state at an earlier time. The two-way influence being proposed here explains the choice of wording "causally symmetric".

---

[2] Note that the letter i is serving double duty here, representing both "initial" and $\sqrt{-1}$.

[3] There is a large literature concerning retrocausality in quantum mechanics, a sample being [5-14].

[4] Note that it is not possible to interpret $\psi_f$ as just a second initial wavefunction to be evolved forwards in time. This is because it must always be an eigenstate of the particular observable yet to be chosen in the future via the experimenter's intervention and this would not be explicable from a forwards-in-time viewpoint. If the experimenter's choice is between two non-commuting observables, there is no single $\psi_f$ which would be compatible with both choices. Also note that $\psi_f$ is not simply accessible as a second initial boundary condition because it will spread in the backwards time direction away from the future measurement event, gradually becoming increasingly entangled in that reverse direction, and any attempt to control it would only change its form in our past, not our future.

## 4. Theory of Measurement

Returning to Eqs. (1) and (2) for the different current densities in the two models, the idea is that the latter provides the current density when both the initial boundary condition $\psi_i$ and the final boundary condition $\psi_f$ are given, whereas the former is just the average current density in the more usual situation where only $\psi_i$ is known. The causally symmetric version also has two different expressions for probability density, depending on whether both the initial and final conditions are specified or whether only the former is given. In formulating a theory of measurement for each model, however, it is only the probability expression given the initial state $\psi_i$ alone which is relevant[5] and this expression turns out to be the same in both models. Hence the argument employed in Bohm's theory of measurement carries through in the same way for either case. Furthermore, this continues to be true in going to many-particle states. An example of this measurement theory in action will now be outlined by way of illustration.

Consider the well-studied situation of a pair of particles in an entangled spin state. This state will be denoted by the spinor wavefunction $\psi_i(\mathbf{x}_1,\mathbf{x}_2;t)$, where $\mathbf{x}_1$ and $\mathbf{x}_2$ refer to the positions of the particles at time t. Spin measurements are to be performed on the 1st and 2nd particle in the directions f and $f'$, respectively. Keeping the analysis fairly general, the possible final outcomes will be described by the spinor eigenfunctions $f_{mn}(\mathbf{x}_1,\mathbf{x}_2;t)$ which represent the joint outcome of the m-th eigenvalue for the 1st particle and the n-th eigenvalue for the 2nd particle $(m,n=1,2,3,...)$. For example, in the simplest case there might be just two possible outcomes for each particle, namely $+\frac{1}{2}\hbar$ and $-\frac{1}{2}\hbar$, and the entangled state might be just the singlet state.

In general the initial two-particle state can be expressed as a superposition of the eigenfunctions $f_{mn}$ as follows:

$$\psi_i(\mathbf{x}_1,\mathbf{x}_2;t)=\sum\nolimits_{m,n} c_{mn}\, f_{mn}(\mathbf{x}_1,\mathbf{x}_2;t) \tag{3}$$

where the $c_{mn}$ are complex coefficients. The aim is to show that the joint probability of obtaining the outcomes m and n is given by $|c_{mn}|^2$. It is assumed here that the eigenfunctions are normalized:

$$\int_{-\infty}^{+\infty} f^{\dagger}_{mn}(\mathbf{x}_1,\mathbf{x}_2;t)\, f_{mn}(\mathbf{x}_1,\mathbf{x}_2;t)\, d^3x_1\, d^3x_2 = 1 \tag{4}$$

---

[5] Discussion of the other expression is therefore postponed to Sec. 8.

where $f^{\dagger}$ is the Hermitian conjugate of f. Now, Bohm's theory of measurement for this situation is based on two assumptions, namely (i) that particles have definite trajectories and (ii) that the positions of the particles are distributed at an initial time $t = 0$ in accordance with the following joint probability density:

$$P(\mathbf{x}_1, \mathbf{x}_2; 0) = \psi_i^{\dagger}(\mathbf{x}_1, \mathbf{x}_2; 0)\, \psi_i(\mathbf{x}_1, \mathbf{x}_2; 0) \qquad (5)$$

The relevant wave equation (e.g., the Pauli equation) then ensures that the form of this distribution continues to hold at later times t:

$$P(\mathbf{x}_1, \mathbf{x}_2; t) = \psi_i^{\dagger}(\mathbf{x}_1, \mathbf{x}_2; t)\, \psi_i(\mathbf{x}_1, \mathbf{x}_2; t) \qquad (6)$$

As already emphasized, expression (6) is common to both models.

Now the measurement process on each particle must allow us to distinguish between the different possible outcomes. For spin measurements this is achieved via magnetic fields which spatially separate the possible results. This separation stage occurs continuously and smoothly. Focussing on the 1$^{st}$ particle as it travels along a definite trajectory, its final alternatives are non-overlapping beams in space and so the particle must flow into just one of these beams. Likewise, the 2$^{nd}$ particle must end up in just one of its spatially separated beams. The measurements are completed by establishing in which beam each particle is located.

Before either of these measurements is carried out, the two particles are most conveniently discussed in terms of a single trajectory in six-dimensional configuration space. Note that it is appropriate to discuss the mathematics of correlated probabilities within this space as long as the physical reality being described remains in three-dimensional space. This point will be pursued further in Sec. 7. Now, in the standard de Broglie-Bohm model, the configuration space trajectory is uniquely determined once the trajectory's initial position is specified, in conjunction with the initial wavefunction. This is no longer the case in the causally symmetric model, where the configuration space trajectory is partially dependent on the experimenter's future choices and hence is different from that in the standard version. The essential point here, however, is that this difference is not relevant in formulating the theory of measurement successfully. It is only necessary that the probability density in Eq. (6) flows with time in the same way for both models.

Turning to the expression for $\psi_i(\mathbf{x}_1, \mathbf{x}_2; t)$ in Eq. (3), the terms in this series become separate packets occupying non-overlapping regions in configuration space. Only one of these terms is consistent with both of the actual outcomes m and n. The joint probability for this pair of outcomes can therefore be obtained simply by calculating the total probability for each particle's **position** to be in the relevant packet. This, in turn, is achieved by integrating the joint probability density $P(\mathbf{x}_1, \mathbf{x}_2; t)$ over the appropriate territory. Now, the wavefunction $\psi_i(\mathbf{x}_1, \mathbf{x}_2; t)$ reduces to $c_{mn} f_{mn}(\mathbf{x}_1, \mathbf{x}_2; t)$ in the relevant region. Hence, from Eq. (6), the joint distribution reduces to $[c_{mn}f_{mn}(\mathbf{x}_1,\mathbf{x}_2;t)]^\dagger c_{mn}f_{mn}(\mathbf{x}_1,\mathbf{x}_2;t)$ in that region. Performing the required integration, the joint probability for m and n is then:

$$\begin{aligned}
P(m,n) &= \int_{\text{mn region}} \psi_i^\dagger(\mathbf{x}_1,\mathbf{x}_2;t)\,\psi_i(\mathbf{x}_1,\mathbf{x}_2;t)\,d^3x_1\,d^3x_2 \\
&= \int_{-\infty}^{+\infty} [c_{mn}f_{mn}(\mathbf{x}_1,\mathbf{x}_2;t)]^\dagger\, c_{mn}f_{mn}(\mathbf{x}_1,\mathbf{x}_2;t)\,d^3x_1\,d^3x_2 \\
&= |c_{mn}|^2 \int_{-\infty}^{+\infty} f_{mn}^\dagger(\mathbf{x}_1,\mathbf{x}_2;t)\,f_{mn}(\mathbf{x}_1,\mathbf{x}_2;t)\,d^3x_1\,d^3x_2 \qquad (7)
\end{aligned}$$

which, using Eq. (4), reduces to:

$$P(m,n) = |c_{mn}|^2 \qquad (8)$$

This is in agreement with the Born rule of quantum mechanics, as required, and hence the correct joint probability distribution for these spin observables has been derived.

This argument can be readily extended to observables other than spin in quantum mechanics and to any number of particles. Hence the Born rule for an arbitrary observable can be derived once the initial joint distribution for the positions of the particles is assumed. Whether or not there might also be final boundary conditions influencing each particle's trajectory is not relevant in the above argument and so the derivation is equally applicable within both the standard and causally symmetric versions of the de Broglie-Bohm model. A final, important point to stress is that both models are on the same footing in terms of needing to assume that the initial joint distribution for position is consistent with the Born rule.

## 5. Some similarities and differences

In comparing the two models, there are certain desirable characteristics they both share. For example, both provide a continuous and smooth description of the measurement process, as opposed to the discontinuous wavefunction collapse of standard quantum theory. Also, both have the advantage of providing a resolution of the well-known measurement problem of quantum mechanics. Standard quantum mechanics is well known to predict that a particle which is not in a definite eigenstate of the measurement to be performed will simply cause the state of the apparatus to become indefinite as well, instead of a successful measurement result being obtained. In both of the models discussed here, however, a definite outcome is obviously achieved from the fact that each particle must finish up inside just one of the spatially separated regions. Wavefunction collapse is then simply the decision to ignore the terms corresponding to other regions in so far as they will have no further physical relevance.

A difference between the two models concerns their versions of determinism. Looking at the single-particle case for convenience, the original model is deterministic once the initial state (including the particle's initial position) is specified. In contrast, the causally symmetric model requires both the initial and final states to be specified (including the particle's position at one instant). It is then deterministic for all intermediate times.

Further differences between the two models will be discussed in the following sections.

**6. Spin: single-particle case**

This section focusses on spin measurements and discusses the corresponding ontological picture which the causally symmetric model implies. After consideration of the single-particle case here, Sec. 7 will examine the entangled two-particle case where the model's consistency with special relativity will be highlighted.

For situations where two successive spin measurements are performed on a particle and the results of both measurements are known, the model will be seen to provide a detailed description of what exists at times between the two measurements. In particular it will give definite values for all of the particle's spin components (most of these remaining hidden, however, because they are not observed). The version of the de Broglie-Bohm model to be presented here adheres to the minimalist viewpoint of Bell and others[6] that the only property possessed by the particle itself is its position as a function of time. Spin is still present, but here it is located in the surrounding regions where both $\psi_i$ and $\psi_f$ are non-zero. It will be seen to exist in the form of a density spread through space in an analogous way to the angular momentum located in a classical electromagnetic field.

Consider a particle which has undergone an initial measurement of spin component i and on which a measurement of the f component is to be performed. Keeping the analysis fairly general, the spin observable corresponding to the f direction will be assumed to have possible eigenvalues $f_n$, with $n = 1, 2, 3, ..$ For example, in the case of a spin ½ particle these eigenvalues would be restricted to the two values $+½\hbar$ and $-½\hbar$. Since the model provides a value for any spin component at intermediate times between the two measurements, it will be convenient to choose an arbitrary further component h and focus on its value. All components other than i and f remain unmeasured. These "unobserved observables" constitute the hidden variables of the model, which explains the choice of the letter h for the extra spin component to be considered.

The model's value for h can be obtained by analogy with Eq. (2) earlier for the current density. The corresponding expression for the spin case would be:

$$\mathrm{Re}\frac{1}{a}\psi_f^\dagger\,\hat{h}\,\psi_i \qquad (9)$$

where $\hat{h}$ is the spin operator for the chosen direction and $\psi_i$ and $\psi_f$ are now spinors. At this point it needs to be noted that $\psi_i$ and $\psi_f$ here are both functions of position $\mathbf{x}$ and so Eq. (9)

---

[6] e.g., Sec. 4 in [15] and Sec. 6 in [16]; see also Sec. 9.7 in [17].

actually describes a spin **density** at each point in space, just as Eq. (2) describes a current density. Hence Eq. (9) indicates that the spin angular momentum is spread around the particle in a field-like manner[7]. Now what is actually needed here is the **total** value for the h component of spin. This value can be obtained by integrating Eq. (9) over all space:

$$h = \mathrm{Re}\int_{-\infty}^{+\infty} \frac{1}{a}\psi_f^{\dagger}\,\hat{h}\,\psi_i\, d^3x \qquad (10)$$

From here on it will be convenient to work in Dirac notation, with the two spinor wavefunctions re-expressed as $\psi_i(\mathbf{x};t) \equiv \langle \mathbf{x};t|i\rangle$ and $\psi_f(\mathbf{x};t) \equiv \langle \mathbf{x};t|f\rangle$. Inserting this notation into Eq. (10), the value of any spin component h, given both an initial state $|i\rangle$ and a subsequent measurement outcome $|f_n\rangle$, is therefore:

$$\begin{aligned} h &= \mathrm{Re}\int_{-\infty}^{+\infty} \frac{1}{a}\langle f_n|\mathbf{x};t\rangle\,\hat{h}\,\langle \mathbf{x};t|i\rangle\, d^3x \\ &= \mathrm{Re}\int_{-\infty}^{+\infty} \frac{1}{\langle f_n|i\rangle}\langle f_n|\mathbf{x};t\rangle\langle \mathbf{x};t|\hat{h}|i\rangle\, d^3x \\ &= \mathrm{Re}\frac{\langle f_n|\hat{h}|i\rangle}{\langle f_n|i\rangle} \end{aligned} \qquad (11)$$

*(There was a misprint in (11) when originally published in Foundations of Physics.)* The h value given by Eq. (11) applies during the intermediate time interval between the i and f measurements. Note that this value has a continuous range and need not be any of the eigenvalues of the h observable. Also note that the integrals in Eqs. (10) and (11) should not be construed as an averaging process. Despite some resemblance to the mean value expressions of quantum mechanics, these equations are defined here to be non-statistical and to yield definite values (i.e., certainties) not averages[8]. The integrations are there simply to convert densities to total values. Finally note that all of this is consistent with the statistical predictions of quantum mechanics because the Bohm theory of measurement discussed in Sec. 4 looks after that aspect.

Mathematical formalism the same as or similar to Eq. (11) has been suggested previously in pursuing the physical reality underlying quantum mechanics, the earliest case known to the present author being Roberts [8] in 1978. It has subsequently been independently discovered and used (with varying physical interpretations) by others, e.g., [11], [12]. In addition, it has been used to predict the mean value of experimental results for weak measurements [18]. The weak value theory also retains the imaginary part of expression (11), treating both parts as experimentally measurable. The stance adopted here goes beyond just describing values generated by measurements to asserting that the equations are also describing the underlying reality existing between measurements. It then seems more appropriate to choose just the real part in Eq. (11), as prompted by the current density expressions in Sec. 2, because it is not clear what meaning could be given to complex angular momentum in real spacetime.

---

[7] For a related interpretation of quantum mechanics having an ontology consisting of densities, see [11] (particularly Sec.3).

[8] In the more usual situation, however, where the future result $f_n$ is not yet known, taking a weighted average over the possible $f_n$ values will then yield the familiar mean value expressions of quantum mechanics.

A curious feature of de Broglie-Bohm models in general is the lack of action/reaction. Although each particle is influenced by its wavefunction, the reverse is not true and the wavefunction is the same regardless of which of the available trajectories the particle is following. Here this feature actually provides a benefit in that the spin value is the same independent of the choice of trajectory and so spin can be discussed without consideration of trajectory details.

The present author has explored the overall spin structure implied by Eq. (11) in the case of a spin ½ particle and confirmed that the resulting spin components in every direction are related to each other via the classical trigonometric rules, in contrast to the peculiar impression conveyed by quantum mechanics that all values are either $+\frac{1}{2}\hbar$ or $-\frac{1}{2}\hbar$. In particular, the following is found for times between two successive measurements performed in different directions on a spin ½ particle when both results are taken to be $+\frac{1}{2}\hbar$:

(i) The value calculated for an arbitrary 3rd spin component h reaches a maximum when its direction is chosen to be in the same plane as the two measurement directions and midway between them.

(ii) This maximum value is given by $\dfrac{\frac{1}{2}\hbar}{\cos\frac{1}{2}\omega}$, where $\omega$ is the angle between the two measured components.

(iii) The value for any other direction at an angle $\theta$ to the maximum direction is equal to $\cos\theta$ times the maximum value.

All of this is consistent with the particle having a single, well-defined spin vector, with components in all other directions being related to this vector via $\cos\theta$, as would be expected physically. When the above prescription for an arbitrary spin component is applied to either of the two directions actually measured, the magnitude reduces to $\frac{1}{2}\hbar$ as required for consistency.

Finally, note that this picture entails the existence of a retrocausal effect in that the direction of the hidden spin vector would be different if a different direction were chosen by the experimenter for the 2nd measurement.

**7. Spin: entangled two-particle case**

In considering entangled states, the main concern is to demonstrate that the underlying physical reality remains consistent with Lorentz invariance despite the constraint arising from Bell's theorem. As a preliminary step, the description provided by standard quantum mechanics will now be summarised in Dirac notation in order to provide the formalism needed for describing the causally symmetric version.

Consider a pair of particles in an entangled spin state. This initial state will be denoted here by an upper case I. Spin measurements are to be performed on the 1st and 2nd particle in directions e and f, respectively, with the possible eigenvalues being $e_m$ and $f_n$ $(m, n = 1, 2, 3, ...)$. Again, these might just be $+½\hbar$ and $-½\hbar$. The reason for choosing the letters e and f will become clearer at Eq. (20) below. The joint probability of getting outcomes $e_m$ and $f_n$, given the initial state I, is given by the Born rule:

$$P\left(e_m, f_n \middle| I\right) = \left|\left\langle e_m, f_n \middle| I \right\rangle\right|^2 \tag{12}$$

The two-particle entangled state $\left|I\right\rangle$ can in general be expressed as a superposition of single-particle states as follows:

$$\left|I\right\rangle = \sum\nolimits_{m,n} c_{mn} \left|e_m\right\rangle \left|f_n\right\rangle \tag{13}$$

where the $c_{mn}$ are complex coefficients. It is understood here that the e kets refer to the 1st particle and the f kets refer to the 2nd. Similarly, the two-particle state $\left\langle e_m, f_n \right|$ on the right hand side of Eq. (12) can be written in terms of single-particle states:

$$\left\langle e_m, f_n \right| = \left\langle e_m \right| \left\langle f_n \right| \tag{14}$$

Pursuing the description of standard quantum mechanics further, suppose for convenience that the measurement on the 1st particle is taken to occur at an earlier time than the measurement on the 2nd particle. In this situation the overall description will be updated at the time of the 1st particle's measurement and the 2nd particle will be assigned a state vector of its own from that time onwards. If the outcome of the 1st particle's measurement is taken to be $e_m$, the updated state for the 2nd particle can be obtained from Eq. (13) by keeping the terms containing $e_m$ and discarding the rest. Since the resulting state is not normalised, a normalisation constant N must also be included. This procedure yields the following single-particle state:

$$\left|i\right\rangle \equiv \frac{1}{N} \sum\nolimits_{n} c_{mn} \left|f_n\right\rangle \tag{15}$$

This is the updated state which quantum mechanics assigns to the 2nd particle once the 1st particle's measurement has occurred, but before the 2nd particle's measurement has been performed. The probability that the 2nd particle's result is subsequently $f_n$, given both the initial entangled state I and the 1st particle's result $e_m$, is then:

$$P\left(f_n \middle| i\right) = \left|\left\langle f_n \middle| i \right\rangle\right|^2 \tag{16}$$

Now this standard formulation raises obvious concerns. If the two measurements are spacelike separated, the description is not Lorentz invariant because the time order of the measurements will differ in different reference frames. Also, the timing of the nonlocal change in the 2nd particle's description from an entangled state to a separate, single-particle state is not Lorentz invariant either, requiring a preferred frame. In contrast, the causally symmetric description avoids these concerns due to its built-in feature, as shown below, that the state $|i\rangle$ can be applied in calculating the properties of the 2nd particle from the moment the initial entangled state is created, not just from the 1st particle's measurement time onwards. Since the form of $|i\rangle$ is dependent on the experimenter's later choice of direction for the 1st particle's measurement, this necessarily entails a retrocausal effect[9]. By this means, the resulting picture for the underlying physical reality avoids the need for spacelike influences.

The mathematical details of this Lorentz invariant description will now be presented. As in the single-particle case discussed in the previous section, it will again be convenient to consider an arbitrary further spin component h, this time its value for the 2nd particle in particular. By extending Eq. (11) to the two-particle case and using the notation of standard quantum mechanics as set out above, the causally symmetric model asserts the following. The 2nd particle's value for an arbitrary spin component h, given the initial state I and the subsequent outcomes $e_m$ and $f_n$, is as follows:

$$h = \mathrm{Re}\frac{\langle e_m, f_n|\hat{h}|I\rangle}{\langle e_m, f_n|I\rangle} \tag{17}$$

where $\hat{h}$ is the corresponding operator for the h spin component of the 2nd particle. Using Eqs. (13) and (14), this expression can be written equivalently as:

$$h = \mathrm{Re}\frac{\langle e_m|\langle f_n|\hat{h}\sum_{r,s} c_{rs}|e_r\rangle|f_s\rangle}{\langle e_m|\langle f_n|\sum_{r',s'} c_{r's'}|e_{r'}\rangle|f_{s'}\rangle} \tag{18}$$

Making use of the fact that the operator $\hat{h}$ only acts on the 2nd particle's state, Eq. (18) can then be rearranged and simplified to:

$$\begin{aligned} h &= \mathrm{Re}\frac{\langle f_n|\hat{h}\sum_{r,s} c_{rs}\langle e_m|e_r\rangle|f_s\rangle}{\langle f_n|\sum_{r',s'} c_{r's'}\langle e_m|e_{r'}\rangle|f_{s'}\rangle} \\ &= \mathrm{Re}\frac{\langle f_n|\hat{h}\sum_{r,s} c_{rs}\,\delta_{m,r}|f_s\rangle}{\langle f_n|\sum_{r',s'} c_{r's'}\,\delta_{m,r'}|f_{s'}\rangle} \\ &= \mathrm{Re}\frac{\langle f_n|\hat{h}\sum_{s} c_{ms}|f_s\rangle}{\langle f_n|\sum_{s'} c_{ms'}|f_{s'}\rangle} \end{aligned} \tag{19}$$

Finally, applying Eq. (15) yields the following value of h for the 2nd particle:

---

[9] As discussed in Sec. 3 of [6], the 2nd particle's state $|i\rangle$ can be thought of as arising at the outset from an inner product of the initial entangled state $|I\rangle$ and the 1st particle's final state $|e_m\rangle$, where the latter extends backwards in time retrocausally from the 1st particle's measurement event to the event where the particles separate.

$$
\begin{aligned}
h &= \mathrm{Re}\frac{\langle f_n|\hat{h}\,N|i\rangle}{\langle f_n|N|i\rangle} \\
&= \mathrm{Re}\frac{\langle f_n|\hat{h}|i\rangle}{\langle f_n|i\rangle}
\end{aligned}
\tag{20}
$$

This expression, however, is seen to be identical to Eq. (11) for the single-particle case! In particular, it is the same as would apply from the outset for an isolated, unentangled particle having initial state $|i\rangle$ and subsequent measurement result $|f_n\rangle$. It certainly does not describe any change in the 2nd particle's state at the time of the 1st particle's measurement. Also, the single-particle nature of Eq. (20) indicates that the ontology of the model is able to avoid the usual six-dimensional configuration space picture for a two-particle entangled system and can provide a picture of independent particles and spin values in three-dimensional physical space. As long as measurements are eventually performed on each entangled particle, it is always possible to derive a separate initial wavefunction for each particle. Furthermore, this reduction to single-particle states can be readily generalised to quantities other than spin and to any number of entangled particles[10]. Although the concept of an entangled state in configuration space is a necessary mathematical tool for correctly calculating the correlated probabilities, all of the physically real ontology for this model resides within spacetime.

## 8. Trajectories

A more contentious difference between the two versions of the de Broglie-Bohm model concerns the way in which their particle trajectories differ. In discussing this point, a clearer perspective is obtained by adopting a relativistic viewpoint using 4-velocity rather than the usual 3-velocity. In the relativistic case, the current densities in Eqs. (1) and (2) both generalise to 4-component expressions, which will here be denoted by $j^\alpha$ $(\alpha = 0,1,2,3)$. Such relativistic expressions for probability flow can in general be expressed in the form:

$$
j^\alpha = \rho_0 u^\alpha \tag{21}
$$

where $\rho_0$ is the probability density in the local rest frame of the flow and $u^\alpha$ is the flow's 4-velocity. In both models the 4-velocity of the particle is taken to be in the same direction as $j^\alpha$ in spacetime and so, since velocity 4-vectors are defined in general to have unit length (assuming units with $c = 1$), the particle's 4-velocity can be equated with the flow 4-velocity and must be equal to $j^\alpha$ divided by its magnitude:

$$
u^\alpha = \frac{j^\alpha}{|j|} \tag{22}
$$

Comparing (21) and (22) then yields:

[10] See Sec. 11 in [5] and Sec. 9 in [6].

$$\rho_0 = |j| \tag{23}$$

in keeping with the fact that the rest density $\rho_0$ must be positive.

Now, the 4-current density of the causally symmetric model for the single-particle case is given by the general expression [6]:

$$j^\alpha = \mathrm{Re}\frac{\psi_f^\dagger\, \hat{j}^\alpha\, \psi_i}{\langle f|i\rangle} \tag{24}$$

where the form of the operator $\hat{j}^\alpha$ depends on the wave equation under consideration. For example, in the Schrodinger case $\hat{j}^\alpha$ has the form:

$$\hat{j}^0 = 1\,,\ \hat{j}^k = \frac{\hbar}{2im}\frac{\overleftrightarrow{\partial}}{\partial x_k} \qquad (k = 1,2,3) \tag{25}$$

At this point the basic assumptions of the causally symmetric model can be stated for the single-particle case. They are that, for an ensemble of independent particles, (i) the initial probability distribution for the positions of the particles, given an initial state $\psi_i(\mathbf{x})$, has the form $\psi_i^\dagger(\mathbf{x})\,\psi_i(\mathbf{x})$, and (ii) the world lines of the particles are distributed in accordance with the 4-current density $j^\alpha$ given by Eq. (24), with each particle's velocity 4-vector lying along the direction indicated by $j^\alpha$. (The generalisation to the entangled, many-particle case is presented in [6].)

From Eq. (24), a notable difference between the two models is now apparent, namely that the time component $j^0$ of the 4-current density is not always positive for the causally symmetric case and so this 4-vector is not always pointing forwards in time. This is a characteristic it shares with the Klein-Gordon 4-current density of standard quantum mechanics. At first sight this may seem a significant difficulty for the model. In the presence of retrocausality, however, this feature can easily be accommodated. In particular, it is straightforward to prove[11] that any such "doubling back" of a world line can only occur at times **between** measurements and so can never be observed. Furthermore, the retrocausal influence automatically ensures that such unconventional behaviour between two successive measurements is gradually smoothed out and eliminated as the time of the second measurement approaches, due to the final wavefunction $\psi_f$ gradually dominating. The world line is then lying within the forwards light cone again, as required. Indeed, by invoking retrocausality one can provide a satisfactory particle interpretation for the standard Klein-Gordon 4-current density as well [19].

---

[11] Appendix 2 of [6]; see also Sec. 6 in [6]. Consistency with special relativity is facilitated and made more transparent by the fact that, although 4-velocity goes to infinity as light speed is approached, the particle's effective 4-momentum remains finite here, thereby providing a continuous description of a smoothly curving world line in spacetime. This is discussed in more detail in Appendix 3 of [6].

The time component $j^0$ of the current density 4-vector is normally interpreted as the probability density for the particle's position. As discussed above, however, Eq. (24) allows the possibility of world line segments curving smoothly through the light cone and temporarily turning backwards in time so that the component $j^0$ is not always positive with respect to our own reference frame. Nevertheless, this is not important because at times between measurements there is no need for this component to provide probability predictions for experiments, its role instead being to describe the direction of the current density 4-vector and the particle's 4-velocity in spacetime. In any case, $j^0$ is always positive in the local rest frame[12] and hence can be interpreted locally as a probability density in that frame.

Although the freedom of a world line to stray outside the forwards light cone is restricted to times between measurements and so is undetectable, its presence in the model may not be to everyone's taste. Nevertheless, it is a necessary trade-off in order to obtain such advantages as avoiding a preferred reference frame and avoiding nonlocal connections in spacetime. Consequently it becomes a matter of personal judgement as to which picture is considered more acceptable.

## 9. Discussion and conclusions

In this paper it has been shown that the Born probability rule for any observable in quantum mechanics can be derived within the causally symmetric version of the de Broglie-Bohm model via an argument analogous to that employed in the original model. Having placed the alternative version on a more equal footing concerning this rule, some of the possible advantages that the model provides have been indicated, such as its ability to avoid a configuration space ontology in the context of entangled states and to maintain consistency with special relativity in the face of Bell's theorem. In particular, it is able to describe entangled particles as being independent, in the sense of no spacelike influences, once retrocausality is included and final boundary conditions are specified.

In discussing these issues, the paper has also highlighted various similarities and differences between the original de Broglie-Bohm model and the present version with regard to basic concepts such as continuity, determinism, time symmetry, underlying trajectories and hidden variables. In terms of ontology, the physically real entities are the particle trajectories plus such familiar quantities as energy, momentum, spin, etc. Wavefunctions are not considered part of the ontology here. Instead they are considered as useful mathematical functions (like, e.g., the six-dimensional Lagrangian in classical mechanics for a pair of mutually interacting particles) from which, once both the initial and final conditions are specified, all of the ontological quantities at intermediate times can be derived.

The present model also demonstrates a possible closer connection between the de Broglie-Bohm picture and the weak value formalism. It essentially merges those two models in a way that incorporates the better features of each. In particular, it combines the de Broglie-Bohm model's ability to resolve the measurement problem and derive the Born rule with the weak value formalism's ability to describe a number of different quantities via the same generic framework and thereby to provide a simple mathematical scheme for incorporating retrocausality.

[12] It is straightforward to extend the concept of a reference frame to superluminal velocities [20]. Also, the continued applicability of Eqs. (21) to (23) in the superluminal case is discussed in Appendix 4 of [6].

Taking a wider perspective, the general intention here has been to present a viable and fully operational example of a retrocausal model in action, in particular one which can encompass all of the quantum mechanical observables. It is hoped that this will facilitate a careful consideration of such models and allow a clear comparison with more common viewpoints[13].

**Acknowledgement**

The author wishes to thank Ken Wharton for his detailed, thorough and valuable feedback on this paper.

**Appendix**

The purpose of this Appendix is to demonstrate that the current density expression of the causally symmetric model as given in Eq. (2) reduces back to the standard current density of the Schrodinger equation, i.e., to Eq. (1), when the final state is not known. These two expressions will be written here as $\mathbf{j}(\mathbf{x};t|i,f)$ and $\mathbf{j}(\mathbf{x};t|i)$, respectively, in order to indicate explicitly whether one or both of the boundary conditions i and f are specified. The aim is to show that the latter expression arises from the former via a weighted average involving the probability $P(f|i)$ of f given i, as follows[14]:

$$\mathbf{j}(\mathbf{x};t|i) = \sum_f \mathbf{j}(\mathbf{x};t|i,f)\,P(f|i) \tag{26}$$

Inserting Eq. (2) for $\mathbf{j}(\mathbf{x};t|i,f)$, Eq. (26) becomes:

$$\mathbf{j}(\mathbf{x};t|i) = \sum_f \mathrm{Re}\frac{\hbar}{2ima}\psi_f^*(\mathbf{x};t)\overleftrightarrow{\nabla}\psi_i(\mathbf{x};t)\,P(f|i) \tag{27}$$

---

[13] An anonymous referee has pointed out a recent paper by Sen [21] which also contains a Bohmian retrocausal approach. Sen's model introduces an ontic $\psi$ determined by future measurement settings in addition to the usual epistemic $\psi$ determined by the initial preparation, with each particle's trajectory determined only by the first of these. By this means, a description is successfully obtained which is local as regards ontology and interactions between particles. It is, however, less time-symmetric than the present model, which treats the initial and final boundary conditions symmetrically and takes them to be on an equal footing causally. Sen's model is able to reproduce the quantum mechanical predictions exactly but, as its author points out, the model as currently presented is restricted to Bell correlations and the retrocausality is assumed in an ad hoc manner rather than defined in physical terms. Sen's paper also refers back to the original presentation of the causally symmetric model in [5] and suggests the criticism that its probability density for position is not always non-negative. However, the updated formulation given here makes it clear that the probabilities predicted for experimental situations are always positive, in accordance with the Born rule.

[14] A discrete spectrum for f is assumed here for simplicity. An analogous proof holds for the continuous case.

Switching to Dirac notation for convenience, the two wavefunctions can be re-expressed as $\psi_i(\mathbf{x};t) \equiv \langle \mathbf{x};t|i\rangle$ and $\psi_f(\mathbf{x};t) \equiv \langle \mathbf{x};t|f\rangle$ and the inner product "a" can be written in the form $\langle f|i\rangle$, so that Eq. (27) becomes:

$$\mathbf{j}(\mathbf{x};t|i) = \mathrm{Re}\sum_f \frac{\hbar}{2im}\frac{\langle f|\mathbf{x};t\rangle \overleftrightarrow{\nabla}\langle \mathbf{x};t|i\rangle}{\langle f|i\rangle} P(f|i) \tag{28}$$

Now one of the basic assumptions of the causally symmetric model is that, given the initial state $\psi_i(\mathbf{x})$, the initial probability distribution for position is $|\psi_i(\mathbf{x})|^2$. This assumption allows the Born rule to be derived for any other observable, as shown in Sec. 4, and therefore allows this rule to be applied for the f observable here. It can be written in the form:

$$P(f|i) = |\langle f|i\rangle|^2 \tag{29}$$

so that Eq. (28) simplifies as follows:

$$\begin{aligned}\mathbf{j}(\mathbf{x};t|i) &= \mathrm{Re}\sum_f \frac{\hbar}{2im}\frac{\langle f|\mathbf{x};t\rangle \overleftrightarrow{\nabla}\langle \mathbf{x};t|i\rangle}{\langle f|i\rangle}|\langle f|i\rangle|^2 \\ &= \mathrm{Re}\sum_f \frac{\hbar}{2im}\langle i|f\rangle\langle f|\mathbf{x};t\rangle \overleftrightarrow{\nabla}\langle \mathbf{x};t|i\rangle \\ &= \mathrm{Re}\frac{\hbar}{2im}\langle i|\mathbf{x};t\rangle \overleftrightarrow{\nabla}\langle \mathbf{x};t|i\rangle \end{aligned} \tag{30}$$

Returning to wavefunction notation and noting that expression (30) is already real without needing to take the real part, this result can finally be expressed as:

$$\mathbf{j}(\mathbf{x};t|i) = \frac{\hbar}{2im}\psi_i^*(\mathbf{x};t)\overleftrightarrow{\nabla}\psi_i(\mathbf{x};t) \tag{31}$$

which is the Schrodinger current density in Eq. (1), as required.